\documentclass
[prl,amsfonts,superscriptaddress,twocolumn,showpacs,floatfix]{revtex4-1}%
\usepackage{amsmath}
\usepackage{amsfonts}
\usepackage{amssymb}
\usepackage{graphicx,graphics,color,epsfig}%

\begin{document}
\preprint{}
\title{Quasiparticles beyond the Fermi liquid and heavy fermion criticality}
\author{Peter W\"{o}lfle}
\affiliation{Institute for Theory of Condensed Matter and Center for Functional
Nanostructures, Karlsruhe Institute of Technology, 76131 Karlsruhe, Germany}
\author{Elihu Abrahams}
\affiliation{Department of Physics and Astronomy, University of California Los Angeles, Los Angeles,
CA 90095}
\date{\today{}}

\begin{abstract}
We give a self-consistent theory of the scale dependent effective mass
enhancement $m^{\ast}/m$ of quasiparticles by 3D antiferromagnetic (AFM) spin
fluctuations in the presence of disorder at an AFM quantum critical point. The coupling of fermionic and
bosonic degrees of freedom in the critical regime is described in terms of
a critical quasiparticle theory. Using the fact that even in the ``non-Fermi
liquid" regime the quasiparticle width does not exceed the quasiparticle energy, we adopt relations
from Fermi liquid theory to determine the dependence of the spin fluctuation
spectrum on $m^{\ast}/m$, from which the self energy and hence $m^{\ast}/m$ may be calculated.
The self-consistent equation for $m^{\ast}/m$ has a strong coupling solution provided
the initial value is sufficiently large. We argue that in YbRh$_{2}$Si$_{2}$ (YRS),
quasi-2D AFM and/or 3D ferromagnetic Gaussian fluctuations existing over a wide range drive the system
into the 3D strongly coupled fluctuation regime. We find critical exponents of
the temperature dependence of the specific heat coefficient $\gamma\propto
T^{-1/4}$ and of the resistivity $\rho(T)=\rho(0)+AT^{3/4}$ in good agreement
with experiments on YbRh$_{2}$Si$_{2}$ in the temperature range $T<0.3K$.

\end{abstract}
\maketitle

\textit{Introduction}. Quantum phase transitions differ from
thermodynamic phase transitions in that temporal fluctuations are of equal
importance or may be even dominant compared to spatial fluctuations. Early
theories of quantum critical behavior, formulated in the framework of a
Ginzburg-Landau-Wilson action of the order parameter field $\phi$ \cite{Hertz,Millis}, found that the effective dimension of the corresponding $\phi^{4}$-field
theory is increased to $d_{eff}=d+z$ where $d,z$ are the spatial
dimension of the fluctuations and the dynamical critical exponent,
respectively. In many cases of interest $d_{eff}$ is above the upper critical
dimension, so that the fluctuations are effectively non-interacting and
the theory is of the Gaussian type.

However, several metallic compounds exhibit a
quantum critical point with non-Gaussian character \cite{HvL,Steglich} (for a
review see \cite{LRVW}). Many of the systems for which the Gaussian theory is
found to fail show strong electronic correlations, as exemplified by a
strongly enhanced effective mass ratio $m^{\ast}/m$ (heavy fermion systems).
One might expect that in such systems it is not possible to integrate
out the fermionic degrees of freedom in favor of a purely bosonic theory.
Rather, in many cases the fermionic properties, in particular the effective quasiparticle
mass, develop critical behavior, which will feed back into the
bosonic spectrum. One is then faced with formulating a critical theory of
coupled fermionic and bosonic degrees of freedom. Moreover, all signs
indicate that such a theory would be in the strong-coupling regime.

Over the past ten years several proposals dealing with this situation have been formulated \cite{Si,pc,SVS}. 
Some of these scenarios have been developed enough to allow comparison with experimentally observed critical exponents, in  particular for CeCu$_{5.9}$Au$_{0.1}$ and for YbRh$_{2}$Si$_{2}$ \cite{CeCuAu,pep}.

In this letter, we analyze the behavior near an AFM QCP within a heavy-quasiparticle picture based on the Anderson lattice model and we refer our results to the heavy fermion compound YbRh$_{2}$Si$_{2}$ (YRS). When the energy scale $T_K$ of the lattice Kondo effect and that of the onset of the critical regime $T_{cr}$ are as wide apart as they are for YRS, namely $T_{K}\approx 25K$ and $T_{cr}\approx 0.3K$, we may assume that well below $T_K$, the Kondo screening and hence the heavy-quasiparticle picture are robust. However, the effect of AFM quantum critical fluctuations leads to a further renormalization of the quasiparticle spectrum, which generates a critical (scaling) behavior of the effective mass \cite{shag}. In contrast to other approaches ({\it e.g.} \cite{pep}), we address the non-Gaussian critical region and consider the interaction of the heavy quasiparticles with AFM fluctuations. We shall show that a self-consistent theory of a
divergent quasiparticle effective mass generated by interaction with 3D AFM fluctuations is capable of providing a sufficiently
accurate account of the critical region as seen experimentally. The interplay of spin fluctuations and fermionic excitations has also been considered in $1/N$ expansion by Abanov and Chubukov \cite{abch} and a renormalization group formulation has been given by Metitski and Sachdev \cite{metsach}.

The above scenario depends sensitively on the detailed nature of spin
fluctuations in a given system. For example, 3D AFM fluctuations
do not lead to true critical behavior, i.e. a Gaussian fluctuation theory is
applicable \cite{Hertz,Millis}, provided the effective mass ratio is not too
large. We argue below that in YRS one has a wide region of
quasi-2D antiferromagnetic (or else 3D ferromagnetic) fluctuations \cite{PW},
which give rise to a substantial enhancement of the effective mass and drive
the system into a strong coupling regime of 3D antiferromagnetic
fluctuations. In YRS, the quantum critical point is accessed by tuning the magnetic field. At the critical value of the field, the crossover from quasi-2D to 3D
antiferromagnetic fluctuations as a function of decreasing temperature takes
place at $T=T_{cr}$. This crossover is clearly seen in experiment. Whereas in
the quasi-2D AFM Gaussian fluctuation regime, marginal Fermi liquid (MFL)  behavior \cite{mfl} is
observed experimentally (by this, we mean specific heat coefficient $\gamma(T)\propto\ln(T_{0}/T)$,
resistivity $\rho(T)-\rho(0)\propto T$), in the critical regime the data are
well-described by different power law dependences: $\gamma(T)\propto T^{\alpha-1}$,
$\rho(T)-\rho(0)\propto T^{\alpha}$. These results follow only when the
strongly anisotropic ($\mathbf{k}$-dependent) contribution of AFM fluctuations
to the quasiparticle energy is smeared over the Fermi surface by sufficiently strong
impurity scattering (elastic mean free path shorter than inelastic mean free path).

\textit{Critical quasiparticle picture.} Our starting point is an Anderson lattice model of correlated $f$-electrons
(spin $\sigma$, site $i$; operators $f_{i\sigma}$) hybridizing with
conduction electrons ($c_{\mathbf{k}\sigma}$), as described by the Hamiltonian%
\begin{align*}
{\cal H}  & =%
{\textstyle\sum\limits_{\mathbf{k},\sigma}}
\epsilon_{\mathbf{k}}c_{\mathbf{k}\sigma}^{+}c_{\mathbf{k}\sigma}+%
{\textstyle\sum\limits_{i,\sigma}}
\epsilon_{f}n_{i\sigma}+U%
{\textstyle\sum\limits_{i}}
n_{i\uparrow}n_{i\downarrow}\\
& +V%
{\textstyle\sum\limits_{\mathbf{k},i,\sigma}}
(e^{i\mathbf{kR}_{i}}c_{\mathbf{k}\sigma}^{+}f_{i\sigma}+h.c.).
\end{align*}

\begin{figure}[h!]
\centering
\includegraphics[totalheight=0.22\textheight, viewport= 120 150 800 560,clip] {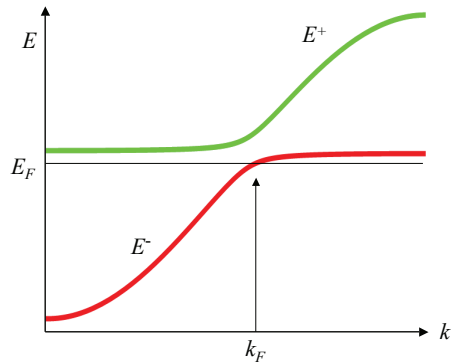}
\caption
{Hybridized bands of the Anderson lattice model. Near half filling, the Fermi level $E_F$ is near the
top of the lower band.}
\label{Figdispersion}
\end{figure}

The single particle Green's functions $G_{ab}(\mathbf{k},\omega)$,
$a,b=\{c,f\}$, may be decomposed into a quasiparticle term and an incoherent
contribution, $G(\mathbf{k},\omega)=zG^{qp}+G^{inc}$, where the quasiparticle
weight factor $z$ is defined by $z^{-1}=1-\partial\operatorname{Re}%
\Sigma(\omega)/\partial\omega|_{\omega=E_{k}}$. Here $\Sigma(\omega)$ is the $f$-electron self
energy generated from the Coulomb repulsion $U$ and which we take to be $\mathbf{k}$-independent. In the limit of large $U$ and close to half-filling ($n_f\lesssim 1$ electrons per
site), one expects $z\ll 1$, leading to a large effective quasiparticle mass
ratio $z^{-1} = m^{\ast}/m$.  We assume that the Fermi level (at $k=k_{F}$) intersects the lower band,
(see Fig.\ \ref{Figdispersion}).
Then the quasiparticle energy is given by
$E_{k}^{-}=$ $(m/m^{\ast})v_{F}(k-k_{F})$, where $v_{F}=2(V/\epsilon_{k_{F}%
})^{2}v_{F}^{0}$ \ is the Fermi velocity of the uncorrelated hybridized band  and the quasiparticle width is 
$\Gamma=z\operatorname{Im}\Sigma(E_{k}^{-})$.
That is to say
$G^{qp}({\bf k},\omega) = [\omega - E_k^- - i\Gamma]^{-1}.
$

The condition for the quasiparticle picture to be valid is $\Gamma<|E_{k}^{-}|$. In the Fermi liquid regime, $\Gamma=c(E_{k}^{-})^{2}\ll |E_{k}^{-}|$ in
the limit $E_{k}^{-}\rightarrow0$. Here, we argue that the
quasiparticle stability condition may be even satisfied in non-Fermi liquid situations.
We extend the usual quasiparticle picture by allowing the parameter $z = m/m^*$ to depend on the energy scale, $z=z(\omega) = 1/[1-\partial\Sigma(\omega)/\partial\omega]$. It is
important to observe that the (retarded) self energy is an analytic function
in the upper half plane, so that the real and imaginary parts of any
nonanalytic term (in the lower half plane) are locally connected. A case of
particular interest is the MFL form of the retarded self
energy at $T=0$ \cite{mfl}: $\Sigma(\omega)= c_{1}[\omega\ln |\omega|/\omega_c - i(\pi/2) |\omega|]$, yielding
the ratio $\Gamma/|E_{k}^{-}|=\pi/(2\ln |\omega_c / E_{k}^{-}|)\ll 1$ .
Thus quasiparticles are still well defined in a MFL. Even in a true non-Fermi
liquid phase with $\Sigma(\omega)\propto -i(i|\omega|)^{\alpha}$,
$\alpha<1$ so that $\operatorname{Im}\Sigma(\omega)\propto
\operatorname{Re}\Sigma(\omega)\propto|\omega|^{\alpha}$ and 
$z\propto(|E_{k}^{-}|)^{1-\alpha}$, one finds $\Gamma/|E_{k}^{-}|=\cot
(\frac{\pi}{2}\alpha)<1$ for $\frac{1}{2}<\alpha<1$. These examples make clear that even if $z=0$ at the Fermi surface, the spectral function for non-zero excitation energy may be
peaked sharply enough to separate a quasiparticle contribution from the incoherent part.

\textit{Spectrum of antiferromagnetic spin fluctuations.} We shall assume that in the heavy-quasiparticle paramagnetic phase, the self energy for the single particle Green's function is determined by the interaction with magnetic fluctuations. Here, we model the
imaginary part of the renormalized retarded dynamical spin susceptibility  for wave vectors ${\bf q}$
near the AFM ordering wave vector ${\bf Q}$ by
\begin{equation}
{\rm Im}\chi({\bf q},\nu)=\frac{(N_0/z)(\nu/v_F^*Q)}{[1+F(Q)+z(\mathbf{q-Q})^{2}\xi_{0}^2]^2 +(\nu
/v_{F}^*Q)^2},
\end{equation}
where $N_{0}$ is the bare density of states at the Fermi surface, $v_{F}%
^{\ast}= (m/m^*)v_F$ is the renormalized quasiparticle Fermi velocity, $\xi_{0}\simeq k_{F}^{-1}$ is
the microscopic AFM correlation length and $F(Q)$ is a dimensionless
generalized Landau parameter, with $F(Q)\rightarrow-1$ at the critical point.
The factor
$z$ multiplying $({\bf q-Q})^{2}$ in Eq.\ (1) arises as follows: Consider the small $({\bf q-Q})$ expansion of the unscreened static quasiparticle susceptibility $\chi \sim (N_0/z)[1 - a({\bf q-Q})^2\xi^2]$. The $({\bf q-Q})^2$ correction is governed by high-energy contributions and should not be renormalized, which leads to $a=z=m/m^*$. Note that
the physical correlation length $\xi=\xi_{0}[(m^{\ast}/m)%
(1+F(Q))]^{-1/2}$ diverges when the critical point is approached in any
direction. As described above, we shall assume $m/m^*=z=z({\cal E})$, where ${\cal E}$
is the relevant energy scale, {\it e.g.} temperature $T$ or magnetic field $H$ (more precisely the Zeeman splitting), whichever is larger.

\textit{Quasiparticle self-energy.} Now we set up a self-consistent determination of the quasiparticle self energy via the leading term in a skeleton graph expansion. Thus, the
quasiparticle width is given by%
\begin{align}
\Gamma(\omega) &=  \lambda^2(Q) \int\frac{d\nu}{\pi}\sum_{{\bf q}} {\rm Im} G^{qp}({\bf k}-{\bf q},\omega - \nu){\rm Im}\chi({\bf q},\nu)\nonumber\\
&\times [b(\nu)-f(\omega-\nu)],
\end{align}
where $\lambda (Q) = zF(Q)/N_0$ is the dimensionful (Landau) interaction vertex and
$f(\omega),b(\omega)$ are Fermi and Bose functions, which at low $T$ confine
the $\nu$-integration to the interval $[0,\omega]$.
This is an equation of self consistency since $\Gamma = z{\rm Im}\Sigma$ appears non-linearly in the integrand.

We consider 3D spin fluctuations and account for impurity scattering by averaging the right hand side of Eq.\ (2) over the Fermi surface. Near the critical line, $F(Q) \approx -1$, and for $Q$ of order $k_F$, we find
\begin{equation}
\Gamma(\omega)=(4/9)(k_{F} \xi_{0})^{-3}(v_{F}Q)^{-1/2}z^{-2}\omega^{3/2}%
\end{equation}
The dependence $\Gamma
(\omega)\propto\omega^{3/2}$ in the case of 3D AFM fluctuations is well known.
The structure of the full self energy $\Sigma(\omega) = \Gamma(\omega)/z$ may be determined from this result as $\Sigma(\omega) \propto (i\omega)^{3/2}/z^3(\omega)$. Here we have generalized to the frequency dependent $z$-factor. We may take a power law form $z(\omega) = b\omega^{\alpha}$
This enables the self-consistent determination
of $z(\omega)= (1 - \partial {\rm Re}\Sigma/\partial\omega)^{-1}$: 
\begin{align}
z^{-1}(\omega)&=1+c'_3(k_{F} \xi_{0}b)^{-3}(v_{F}Q)^{-1/2}3(1/2-\alpha)\omega^{1/2-3\alpha}\nonumber\\
&= 1+c_3(k_{F} \xi_{0})^{-3}(v_{F}Q)^{-1/2}z^{-3}(\omega)\omega^{1/2}
\end{align}
where $c_3$ is O(1). 
We explore the consequences of the scale dependent $z$. For frequencies less than the temperature, we replace $\omega$ by $T$ . Since
$c_3(k_{F} \xi_{0})^{-3}\approx 1$, we can say that as long as
$z^{-3}(T)(T/v_{F}Q)^{1/2}\ll 1$ for any $T$, the system will be in the
Gaussian fluctuation regime all the way down to the critical point. If
however, the initial value of $z^{-1}(T)$, when one enters the 3D AFM
fluctuation regime, is sufficiently large, such that $z^{-3}(T)(T/v_{F}%
Q)^{1/2}\gg 1$, a new regime is accessed, which is of a strong-coupling nature.
We find the characteristics of this new regime within the present
approximation by solving the self-consistent Eq.\ (4), to get
\begin{equation}
z(T)=[c_{3}(k_{F}\xi_{0})^{-3}]^{1/2}(T/v_{F}Q)^{1/4}.%
\end{equation}
In the case of only 3D AFM fluctuations it is difficult to satisfy
the strong-coupling condition unless $z^{-1}(T)$ is sufficiently large. Therefore, if on the initial approach to the
critical point, fluctuations dominate that lead to a growing $z^{-1}(T)$ with
decreasing $T$, the condition may be met at some point. The precise crossover
point is determined by the crossover of these precursor fluctuations to
the critical 3D AFM fluctuations and by the condition above that leads to Eq.\ (5). As mentioned
in the introduction, there are clear indications in the data on YbRh$_{2}%
$Si$_{2}$ of both quasi-2D AFM and 3D FM fluctuations. In both cases one finds
$z^{-1}(T)\propto\ln(T_{0}/T)$, so that $z^{-1}$ grows as $T\rightarrow0$ and is about 40 in the
heavy Fermi liquid region of the phase diagram.

\begin{figure}[h]
\centering
\includegraphics[totalheight=0.22\textheight, viewport= 70 130 800 560,clip] {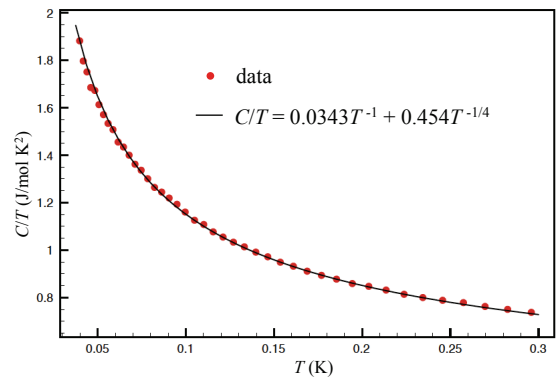}
\caption
{Specific heat: Comparison of theory, Eq.\ (6) and data of Ref.\ \cite{oechs} at the critical magnetic field and below the critical temperature for quantum critical scaling.}
\label{gamma}
\end{figure}

\textit{Specific heat and electrical resistivity.} Within the approximation of
neglecting the momentum dependence of the self-energy, the entropy density
is given by $S/V=(2N_{0}/T)\int d\omega\,\omega(-\partial f/\partial
\omega)[\omega-\operatorname{Re}\Sigma(\omega)]$. Substituting the power law
dependence found above, $\operatorname{Re}\Sigma(\omega)\propto|\omega
|^{3/4}\,{\rm sign}(\omega)$, we find a specific heat coefficient diverging in the
limit $T\rightarrow0$,%
\begin{equation}
\gamma(T)=c_{\gamma}N_{0}(T/v_{F}Q)^{-1/4}%
\end{equation}
A comparison of the theoretical temperature dependence of Eq.\ (6) with experiment \cite{oechs} is shown in Fig.\ \ref{gamma}. To achieve this excellent fit, a $T$-independent constant specific heat has to be added. Such a term could arise from very low frequency ($\omega \ll T$) oscillators. Its magnitude represents about 0.4\% of the total number of formula units. A possible source of such low frequency oscillators is spatially and temporally fluctuating AFM domains that oscillate about the preferred $c$-direction. An anisotropic exchange interaction has been proposed to explain the observed electron spin resonance $g$-shift \cite{WA}. We estimate the corresponding oscillator quantum as $\hbar \omega \approx 0.5 \sigma$ K, where $\sigma \ll 1$ is the staggered magnetization per formula unit of a typical domain.

As mentioned above, beyond the critical regime proper, at temperatures
$T>T_{cr}\approx0.3K$, the data on YRS indicate the existence of
Gaussian fluctuations of quasi-2D AFM or 3D FM character. This leads to
$\gamma(T)=c_{G}N_{0}(\partial/\partial T)[T\ln(T_{h}/(T+T_{FL}))$, where
$T_{FL}$ is the crossover temperature into the Fermi liquid regime and
$T_{h}\approx20K$ is a high temperature cutoff scale of the order of the
lattice Kondo temperature.

\begin{figure}[h]
\centering
\includegraphics[totalheight=0.22\textheight, viewport= 20 100 800 560,clip] {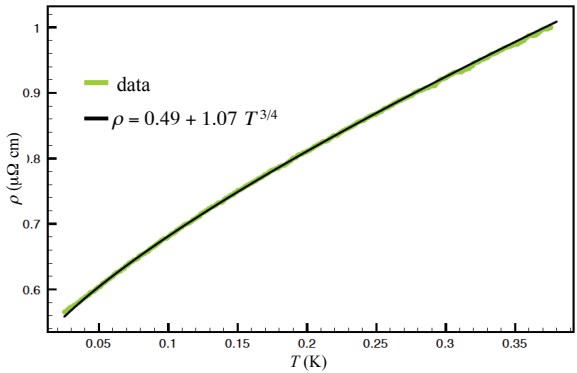}
\caption
{Resistivity: Comparison of theory, Eq.\ (7) and data of Ref.\ \cite{geg} at the critical magnetic field and below the critical temperature for quantum critical scaling.}
\label{res}
\end{figure}

The electrical resistivity in the presence of impurity scattering may be
obtained from the quasiparticle relaxation rate $\Gamma$ as $\rho
(T)=\rho(0)+c'_{\rho}(m/e^{2}n)(m^{\ast}/m)\Gamma$ . Using the above results in
the scaling regime we find%
\begin{equation}
\rho(T)-\rho(0)=c_{\rho}(m/e^{2}n)(v_{F}Q)^{1/4}T^{3/4}%
\end{equation}
Upon entering the Gaussian fluctuation regime at $T>T_{cr}$ this fractional
power law behavior crosses over into a linear $T$-dependence. A comparison of the
theoretical temperature dependence of Eq.\ (7) with experiment \cite{geg} at the critical magnetic field
is shown in Fig.\ \ref{res}.

\textit{Local susceptibility. }The local ($q$-integrated) susceptibility
$\chi_{loc}(\omega)$ determines the nuclear spin relaxation time $T_{1}$
through the relation%
\begin{align*}
\frac{1}{T_{1}T}  & \propto[\frac{1}{\omega}\operatorname{Im}\chi_{loc}%
(\omega)]_{\omega\simeq 0} =\sum_{{\bf q}}\frac{1}{v_{F}Q\xi_{0}^{4}}\frac{N_{0}(m^{\ast} /m)^{2}%
}{[\xi^{-2}+q^{2}]^{2}}\\
& =N_0(m^*/m)^4(\xi/\xi_{0})/(8\pi v_F Q\xi_0^3).
\end{align*}
where $\xi$ is the physical spin correlation length already defined below Eq.\ (1) as $ \xi = \xi_{0}[(m^{\ast}/m)%
(1+F(Q))]^{-1/2}$ and we used Eq.\ (1). Close to,
but not quite at the critical point, such that $\lim_{T\rightarrow
0}(1+F(Q))>0,$ we then find 
\begin{equation}
\frac{1}{T_{1}T}\propto(\frac{m^{\ast}}{m}%
)^{7/2}= z^{-7/2}\sim T^{-7/8}.
\end{equation}
Exactly at the critical point the temperature
dependence of the control parameter enters, $[1+F(Q)]\propto T^{\beta}$, and
$1/T_{1}T\propto T^{-7/8-\beta/2}$; the value of $\beta$ is not
known at present. Eq.\ (8) is a prediction of our theory; unfortunately the lowest field used in the available data \cite{ishida}, 0.15 T, is about twice the critical field of 0.06 T and is thus outside the critical region

\textit{Conclusion.} We have presented a critical quasiparticle
theory of heavy fermion compounds near a critical point. We focused on
YbRh$_{2}$Si$_{2}$, but the general framework of our approach should be
applicable to other systems. The starting point is the
observation that quasiparticles are well-defined in certain classes of
non-Fermi liquid states. The best known example is the marginal Fermi liquid, properties of which are
observed in a certain region of the $T-H$ phase diagram of YRS. Close to the critical point, however, much stronger deviations
from Fermi liquid theory are observed. We argue that the MFL behavior is caused by Gaussian fluctuations of quasi-2D antiferromagnetic and/or 3D
ferromagnetic character. These fluctuations lead to a quasiparticle effective mass
$m^{\ast}$ increasing logarithmically with temperature. By contrast, 3D
antiferromagnetic fluctuations do not lead to an increasing $m^{\ast},$ unless $m^{\ast}$ exceeds a certain threshold. Then a strong coupling regime is
reached. We calculated the critical exponents in that regime in
a skeleton graph approach.  Results for specific heat, resistivity and NMR relaxation rate are in Eqs.\ (6-8). In Figs.\ 2,3, we compare our results with
the low-$T$ data on specific heat and resistivity and find reasonable agreement. 

We thank A.V. Balatsky, M. Graf, I. Martin, A. Rosch, Q. Si, C.M. Varma, and M. Vojta for useful
discussions and S. Friedemann and N. Oeschler for sharing data. This work was supported in part by the
DFG research unit 960 ``Quantum phase transitions." Part
of this work was carried out at Los Alamos
National Laboratory (PW) and the Aspen Center for Physics (EA,PW).

\end{document}